# Twisting Rolls.
# An Heuristic Model and 3D Numerical Simulations of Vortex Patterns

**Safieddine Bouali**
*University of Tunis, Institut Supérieur de Gestion*
*41, rue de la Liberté, 2000 Le Bardo, Tunisia*
*Safieddine.Bouali@isg.rnu.tn*

(June 2003)

**Abstract**:

We connect an appropriate feedback loop (i.e. the twist process) to a model of 2D vertical eddy of airflow which unfolds a wide range of vorticity behavior. Computational fluid dynamics of the twisted roll display a class of long lifespan 3D vortices. On the one hand, the infinitely stable columnar vortex simulated describes waterspouts and tornadoes with extended lifetime. On the other hand, a light modification of the retroaction exhibits a singular topology of the vortex since the flux moves on to a quasi-torus with a *wind up* surface (spiralling curvature). This phenomenology have strong similarities to the tropical cyclones. Besides, we detect the *proto-cyclone* structure when the control parameter of the retroaction is very low.
Moreover, we investigate the outcome of the twisting process vertically shifted. This modelisation leads to the simulation of simultaneous vortices similar to the severe tornadoes among which one looks like the small suction whirl (the microbrust). The dependency of initial conditions and parameters is associated to this other class of 3D vortices with short lifespan.
Our heuristic dynamical systems lay the foundations of a comprehensive modelisation of vortices since it joins theory and numerical simulations.

# 1. Introduction

The exploration of the vortex genesis constitutes topics and aims of a wide literature in several fields of physical researchs (Atmospheric and Oceanic Sciences, Geophysics, Astrophysics, Superconductivity, Mathematical Physics, Nonlinear Physics...). However, the prominent contributions to the vorticity patterns analysis [see Saffman, 1992; Chorin, 1994 and Benkadda & Zaslavsky, 1998] apply the standard tools of fluid mechanics.
Atmospheric sciences [Bengtsson & Lighthill, 1982; Pedlosky, 1987; Peixoto & Oort, 1992] for example use fluid dynamics models (Newton, Euler and Navier-Stokes equations, boussinesq's approximation...) its variables and parameters (viscosity, compressibility, Rossby and Reynolds numbers,...) to vortex studies.
Our research is focused on the computational fluid dynamics of vortices and whirlwinds in the small, meso and large-scale atmospheres. Nevertheless, we select from the fluid characteristics only those kinematics which are well modelised by autonomous systems of ordinary differential equations (O.D.E.) in the Euclidian plane or space.
Our paper considers the feedback loops modifying the 2D roll system – rotating convection or eddy- the spring of the vortex genesis. Indeed, this phenomenon appears as an outcome of a retroactive (and intricate) connection between the three dimensions of the fluid flow.
In previous papers [Bouali, 1999 & 2002], we explored the results of this methodology into a periodic system. It establishes a full feedback linkage and leads to the formulation of a new system. The retroactive loop injects anti-equilibria excitation which have shown the power of this "endogenous" forcing thereby building an utterly different dynamics. One would expect a similar behavior in the present contribution.
In *sec. 2*, we propose a nonlinear 2D model as a dynamical analogue of any (vertical) eddy and introduce its characteristics.
In *sec. 3*, we connect appropriate retroactive linkages to the 2D model determined by the state variables of the roll itself. We identify the axisymmetric vorticity patterns and analyse briefly the general stability of the new systems. Our modelisation is based on an heuristic approach of the vortex genesis as it is a tool to explore the findings of the feedback linkages. In this direction, we examine, in *sec. 4*, the numerical outcome of two embedded linkages into our basic 2D model of roll.

Our main objective is the universal modelisation of vorticity in physical fields but we kept the atmospheric observations (tornadoes, hurricanes...) as our fundamental reference of vortex structures.
Our modelisation is largely a qualitative application of nonlinear fluid mechanics and the results for vortical patterns are qualitative to a large extend.
Eventually, in *sec. 5*, we point chiefly at the next steps of this framework. We notice that the systems are numerically integrated using the 5$^{th}$ order Runge-Kutta method with $10^5$ unit time.

## 2. The 2D numerical analogue of a roll

The first step of our vortex modelisation requires the choice of a convection roll formula. The governing equation of the 2D periodic motion should emulate the pure kinematic properties of a rotating thermal convection [Emanuel, 1994] or an eddy created by (directionally or speed) wind shear.
We define the system of two ordinary differential equations:

$$\begin{cases} dx/dt = -x(a - by) \\ dy/dt = y(c - x^2) \end{cases}$$

where x and y are the coordinates of one air parcel in the vertical plane and a, b and c positive parameters.. We use in this work the Euclidian system *XY*, so that the vertical axis is *Y* and *X* the horizontal direction. Our set of two first-order coupled equations constitutes a mathematical analogue of a generic vertical roll of fluid and estimates the velocity of the flux. Our 2D dynamical system is derived from the Gause equation [see Appendix].
The first item of the horizontal velocity, i.e. – *ax*, represents the fluid friction. The second one, i.e. *bxy*, indicates the nonlinear diffusion of the flux. Besides, it accelerates the motion for the high levels of Y.
On the other hand, the first item of the vertical velocity, i.e. *cy*, stands for the diffusion and the second one: -*x²y* is the acute nonlinear braking of the flow reinforced when the flow is far from the origin x = 0. Both velocities are not identical over the roll and no symmetry of the upward and downward fluxes is found since the fall of the flow is accelerated by the low pressure of weak altitudes. Besides, the rotor created by wind shear have a strong downward motion since the gravity injects better diffusion to the air parcels.
Associated with the fluid mechanics, the three parameters allow strong plasticity of the roll sketch. Indeed, the variation of *a* and *c* parameters imply a scale modification of the roll, while arising *b* the orbit is stretched in the vertical direction.
In fact, the eddy is provided with a wide spectrum of morphological plasticity allowing to the path a conformation ability to the numerous constraints (orographic topography...). The dynamics replace the *solid wall* boundaries conditions used in many convection researchs.
The numerical computation gives a clockwise periodic orbit (Fig. 1) for initial conditions in the $X^+Y^+$-plane.

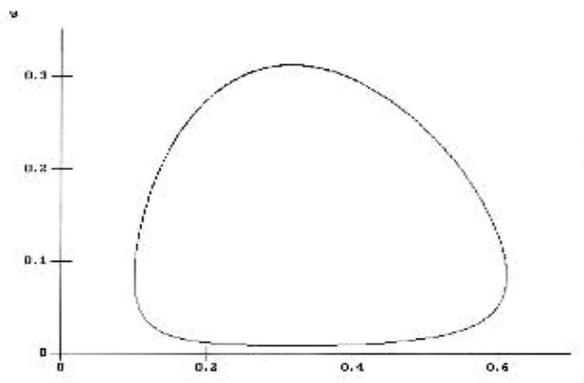

*Fig. 1. The roll sketch.*
*$P_1$ (a, b, c) = (0.05, 0.6, 0.1) and Initial Conditions IC ($x_0$, $y_0$) = (0.1, 0.1).*

Nonetheless, for several initial conditions, the orbits are concentric and centred around a stable equilibrium. For the parameters $P_1$ (a, b, c), the stable solution is $S_1$ (x, y) = ($c^{0.5}$, a/b). Moreover, with a symmetric initial conditions, i.e. in the $X^-Y^+$-plane, the roll follows a counterclockwise periodic orbit with identical characteristics of concentric orbits around another stable solution $S_2$ (x, y) = (-$c^{0.5}$, a/b). Eventually, for the trivial solution $S_0$ (x, y) = (0, 0) the two real eigenvalues are opposite and it is a saddle-node.
For realistic flows, it is to note that there are some example which do not fall in our dynamical system but our objective is building a stylised framework with simple but exemplary model of rotational fluid.

This paper explores the outcome of the advection feedback which inserts the twist process onto the vertical roll according the Z-axis.

## 3. The class of long lifespan vortices

Explaining the ability of fluid dynamics to display persistent vortex organisations remains an exciting field of research and a challenging topic.
We investigate the outcome of the twist process applied to the vertical eddy according the advection channel It is assumed that the feedback function connecting the numerical roll with additional (and linear) ordinary differential equation founds a new 3D system.
Firstly, we explore numerically the flow patterns when the equation governing the advection flow (z-equation) has a simplest definition.
Secondly, a system with modified twist equation is investigated. In both cases, we attempt to detect similarities of the simulated and idealised dynamics of our heuristic approach with the vortical structures described in the literature of atmospheric sciences.

### 3.1. Elementary vortex dynamics

We connect to the roll formula an additional equation leading to a 3D system with advection linkage:

$$\begin{cases} dx/dt = -x(a - by) + sz \\ dy/dt = y(c - x^2) \\ dz/dt = -dx \end{cases}$$

where *XZ* is the horizontal plane (or ground level). The z-equation determined by the state variable x of the previous 2D model itself indicates the inertia term of the advection flux and *d* stands for its braked horizontal velocity. This feedback loop controlled by the *s* parameter founds a new system.
With positive parameters $P_2$ (a, b, c, d, s), the real part of an eigenvalue is negative. It is adequate to qualify the instability of the trivial solution $S_0$ ( x, y, z) = (0, 0, 0). However, computations are required to qualify the patterns of the dynamics.
The simulations display a columnar vortex at *the hub* of an annulus torus (fig. 2) with upright whirlwind from the initial conditions. At the top, the flux moves downwards over the torus then follows a convergent spiral on to the *XZ*-plane but before reaching the equilibrium, it arises in a second whirlwind along a different path in the same funnel. The axisymmetric vortex builds a closed domain where the inflow-outflow flux constitues the surface of the annulus torus.
In accordance with these simulations "...a *strong radial inflow near the surface in tornadic vortices [...] approaches the central axis of the vortex, the radial inflow separates from the surface, turns upward, and flows up and out of the vortex core*" [Nolan & Farrel, 1999, p. 1283].

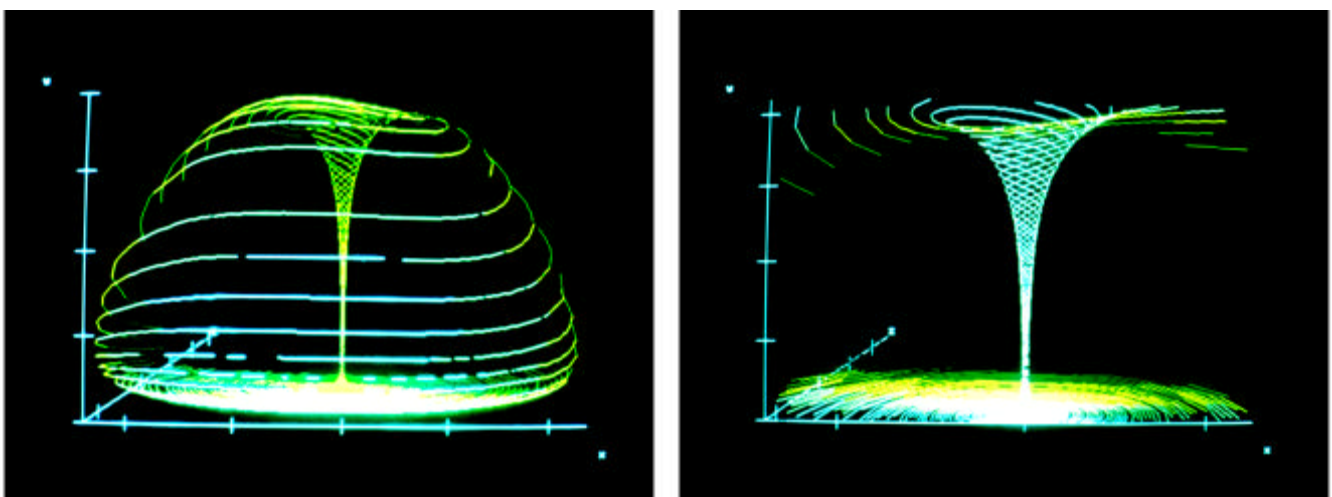

*(a) External view of the annulus torus*  *(b) Inside the torus*

*Fig. 2. Waterspout-like vortex dynamics.*
*$P_2$ (a, b, c, d, s) = (0.2, 0.6, 0.1, 10, 0.5) and IC ($x_0$, $y_0$, $z_0$) = (0.1, 0.01, 0.01). The surface level inflow spirals toward the axis of the torus, it swirls upwards and constitutes the outflow at the apex of this long lifespan vortex.*

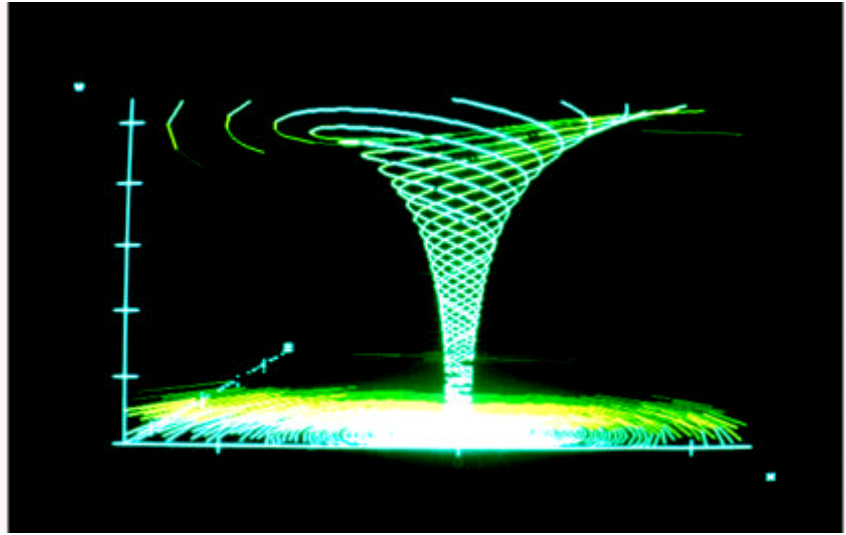

*Fig. 2.1. Axisymmetric tornado-like dynamics.*
*$P_{2.1}$ (a, b, c, d, s) = ( 0.2, 0.6, **0.2**, 10, 0.5 ) and IC ($x_0$, $y_0$, $z_0$) = (0.1, 0.01, 0.01). A small modification of a parameter widens the columnar vortex.*

We notice the non-stop numerical computations since the iterates do not converge to a null state and testify the long lifespan of these vortices. The kinematics of the twisted convection produce an incessant stable vorticity and its robust dynamics may be broken only by a modification of the 3D system itself.

If this is the case, then it will be clear that dynamical *lock-in*, i.e. eddy twisted by an advection flow, plays a major role in the vorticity genesis. The whirlwinds depicted reproduce well enough the phenomenology of spray vortex (waterspouts) reported in the literature [Gordon & Purcell, 1978; Brady & Szoke, 1989; Simpson *et al.*, 1991; Lin, 92] and the long lifetime tornadoes (Fig. 2.1).

Otherwise, singular and stable whirlwinds [Wood, 1992] occur when strong rotational thermal convection is associated with light wind. These particular extended lifespan vortices are well described by our twisting roll perspective. We derive from this *dynamical setting* of fluxes a version with modified retroactive loop suited also to an heuristic approach.

### 3.2. Vortex on to a *wind up* surface

We explore the characteristics of the new 3D system:

$$\begin{cases} dx/dt = -x(a - by) + sz \\ dy/dt = y(c - x^2) \\ dz/dt = -d(x + y) \end{cases}$$

when the feedback is composed of a linear sum of the *XY* coordinates. It leads us to the detection of the unstable steady state $S_0$ (x, y, z) = (0, 0, 0) since the determinant of the Jacobean is positive with a set of positive parameters $P_3$ (a, b, c, d, s). The nontrivial solution $S_1$ (x, y, z) = (-$c^{0.5}$, $c^{0.5}$, $c^{0.5}$ (b.$c^{0.5}$ - a)/s) is also unstable, but we reject the second solution $S_2$ since the coordinate of *Y* is negative.

In order to study the dynamics of the system, it plotted a Poincaré map (Fig. 3.a) monitoring the impacts of the trajectory of flow with a vertical plane cutting the centre of the annulus torus. Contrarily to the previous torus, it is not a closed and hermetic domain. The trajectory moves on to a rolling up (and spiralling) surface which builds the torus itself (Fig. 3.b). In spite of its basic formulation, a more intricated topology of the trajectory path is produced. Indeed, the first whirl rushes in the vertical direction and the same flux designs the second and concentric eyewall (Fig. 3.c), then the hidden third, fourth...eyewalls.

The overlapped swirling flows *are* the vortex and the upward and downward winds build the wrapped curvature of the torus. From the initial conditions, the whirlwind flow arises in the eyewall (*it is the first eyewall !*) then sinks following a gradient trajectory over the peripheral side of the torus and repeat its dynamics inside the torus itself. Eventually, the flux will be trapped into a stable orbit. This horizontal roll (Fig. 3.d) crosses the two points in the centres of the two spirals in the discrete two-dimensional map. The model dives the flux into an infinitely (in the sense of numerical studies) fixed structure.

This representation does not contradict the studies of hurricane structure [Lilly, 1986; Bartels & Maddox, 1991; Montgomery & Farrel, 1993] with explicit confirmation of the non-unique eyewall [NOAA, 1999].

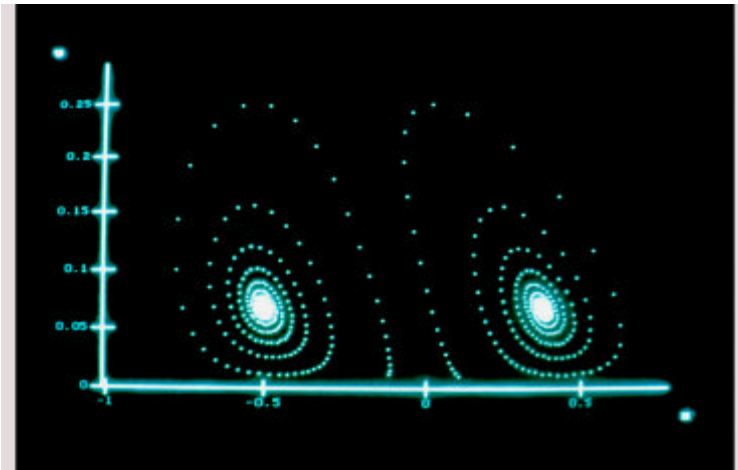
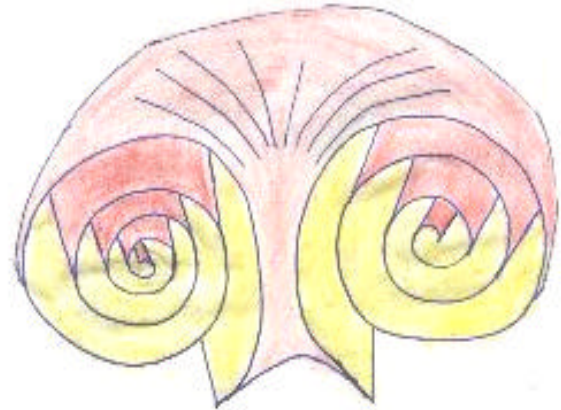

*(a) A Poincaré map of the quasi-torus*  *(b) The quasi-torus split*

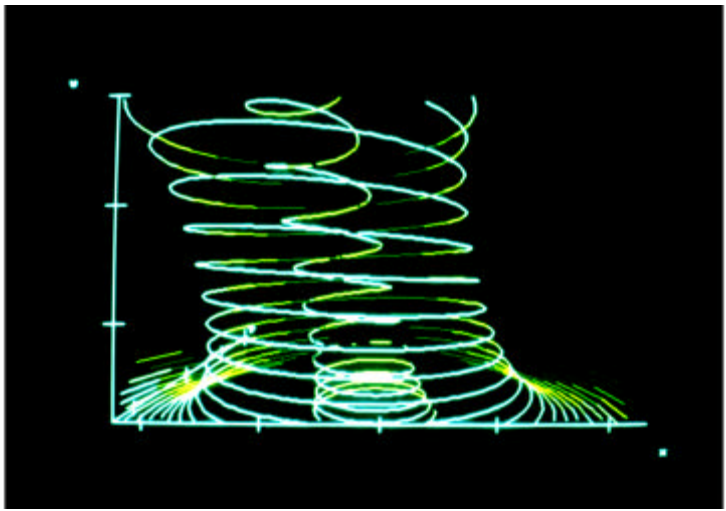
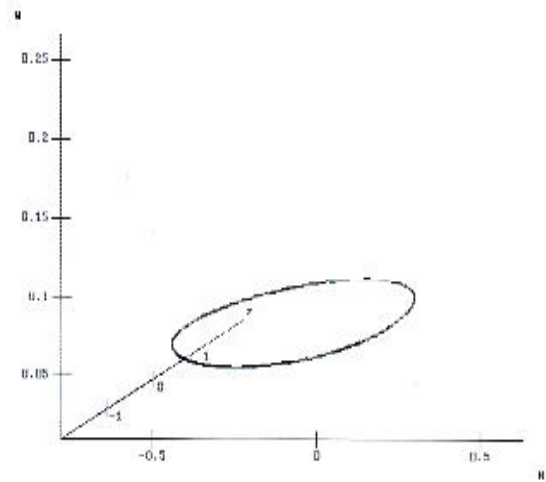

*(c) The overlapped whirlwinds*  *(d) The final periodic motion*

*Fig. 3. Wind up vortex dynamics.*
$P_3 (a, b, c, d, s) = (0.05, 0.6, 0.1, 10, 0.5)$ and IC $(x_0, y_0, z_0) = (0.1, 0.01, 0.01)$

Moreover, the vorticity stretching of the hurricanes is determined *previously* by the extension of the roll shape in the 2D vertical plane. The eddy size appears as a variable in the track of atmospheric vortices since the vertically compressed roll will modify the aerodynamics of the expected annulus torus and will be insensible to the prevalent wind.

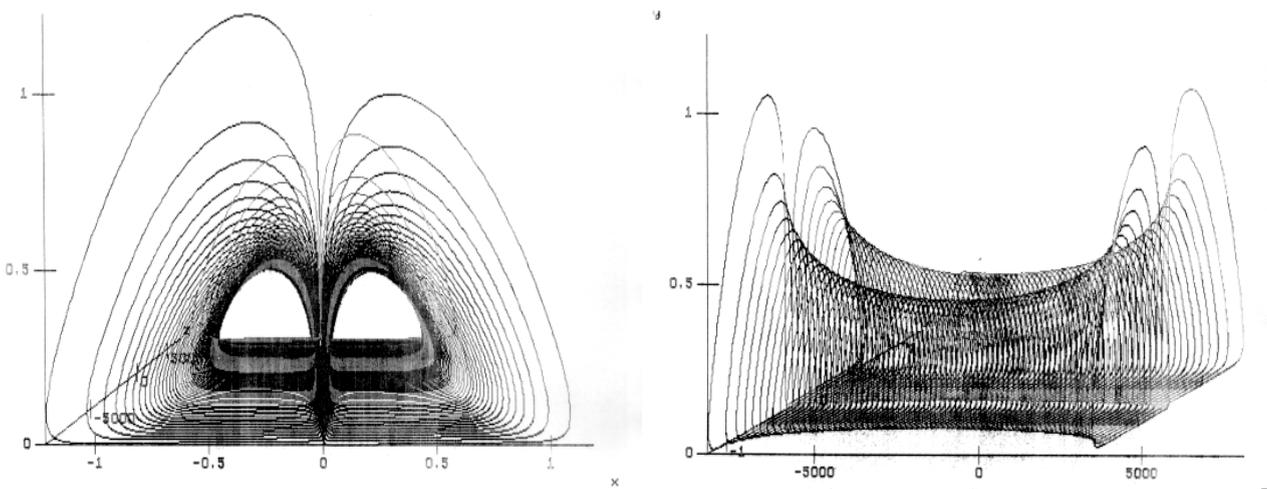

*Fig. 3.1. (a) and (b) two horizontal views of the Proto-Cyclone.*
$P_{3.1} (a, b, c, d, s) = (0.05, 0.6, 0.1, 10, 10^{-8})$ and IC $(x_0, y_0, z_0) = (0.1, 0.01, 0.01)$.

Besides, the numerical computations predicts a singular pattern of the fluid mass representing the precursor state immediately preceding the cyclonic vortex. This *proto-cyclone structure* (Fig 3.1) occurs when the control parameter of the retroaction is very low (s = $10^{-8}$). In this case, the airflow builds two *solenoidal* rolls with clockwise and anticlockwise paths. This simulation suggests that this pattern seems to be an *immature* stage of the cyclone between the tropical wave disturbance in the atlantic observed prior the cyclone warning and the fully developed hurricane structure.

In this section, all the simulations have an infinite horizon and the above description of the long lifespan vortex indicates the conservation of the phase space volume. The present computational framework leads our heuristic approach of the fluid dynamics to the next numerical experiments when the vortex patterns vanish after a finite number of iterates. Indeed, we define and link a supplemental retroaction to the twist process.

## 4. The class of short lifespan vortices

Our perspective is heuristic since the modelisation serves as a tool to know the outcome of sophisticated linkages into the 2D roll. In order to understand the powerful generation of this twist process, it is enlightening to examine the effect of two embedded feedbacks. This amendment modifies the previous model as we connect another feedback loop to the first retroaction. Now, the flux is governed by this extended model:

$$\begin{cases} dx/dt = -x(a-by) + sz \\ dy/dt = y(c-x^2) \\ dz/dt = -dx + vw \\ dw/dt = -fy \end{cases}$$

which contains an additional control parameter v allowing the *fine-tuning* of the w-variable. A new 4D system is built and the fourth dimension performs a vertical shift *distorting* the twist process itself according to the f parameter.

A brief general stability exhibits two stable equilibria. For the positive parameters $P_4$ (a, b, c, d, v, f, s), the first solution is $S_1$ ( x, y, z, w) = ($c^{0.5}$, 0, $a.c^{0.5}$/s, $d.c^{0.5}$/v) and the second solution is $S_2$ ( x, y, z, w) = (-$c^{0.5}$, 0, -$a.c^{0.5}$/s, -$d.c^{0.5}$/v).

Several vortical patterns associated to a strong Sensitive Dependency on Initial Conditions (SDIC) appear from the numerical explorations. Indeed, simultaneous vortices are simulated (Fig. 4) among which one is provided with high swirling speed displays the bursts phenomena reported earlier [Fujita, 1981; Forbes & Wakimoto, 1983; Hooze, 1993]. This tiny suction vortex (Fig. 4b) gushes from the ground and follows a faster *ballistic* motion in the decaying stage of the simulated system which stops after a small number of iterates. It testifies the lifespan brevity of this second class of 3D vortices.

Obviously, our numerical simulations look like the severe tornadoes [Rasmussen *et al.*, 1994] and the smaller sucking whirls called micro- and downbursts phenomena.

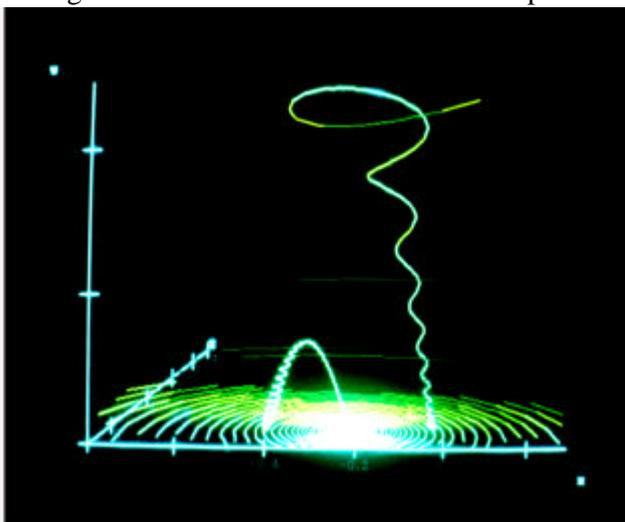
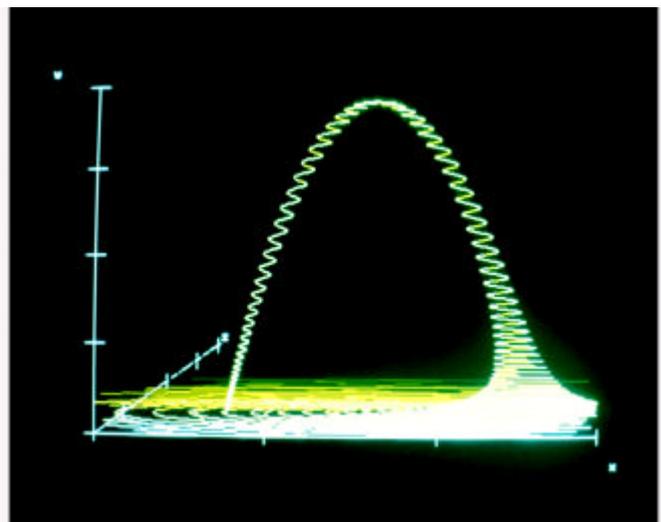

*(a) The mono-column vortex with a ballistic swirl.*  *(b) The ballistic swirl (zoom).*
 $IC(x_0, y_0, z_0, w_0)$= (0.01, 0.01, 0.01, 0)   $IC(x_0, y_0, z_0, w_0)$= (0.01, 0.01, 0.01, 0)

**Fig. 4. The Sensitive Dependency on Initial Conditions (SDIC)**
*$P_4$ (a, b, c, d, v, f, s) = ( 0.05, 0.6, 0.1, 5, 0.05, 1, 0.5 )*

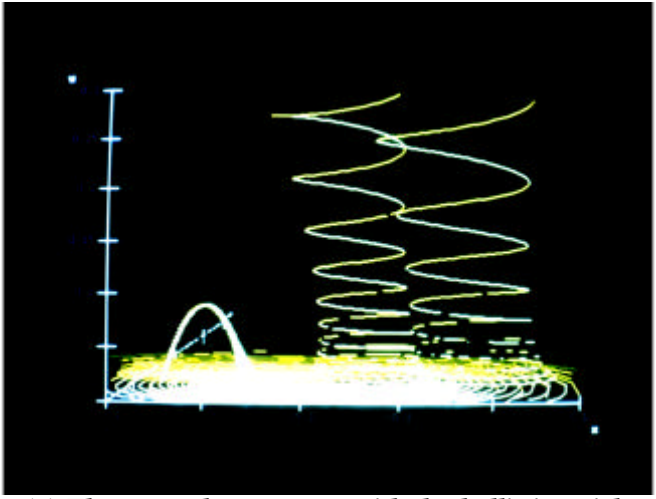
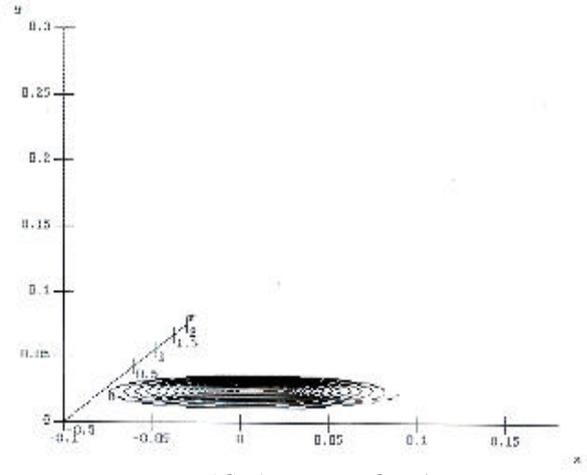

*(c) The two-column vortex with the ballistic swirl.*
$IC(x_0, y_0, z_0, w_0)= (0.1, 0.01, 0.01, 0)$.

*(d) A vortex abortion.*
$IC(x_0, y_0, z_0, w_0)= (0.1, 0, 0, 0)$.

***Fig. 4.** (continued)*

Indeed, when the tornado as "a violent rotating column of air in contact with the ground", literature (and chase teams news!) describes also multiples main funnels in tornado dynamics [Gall, 1985; Church *et al.*, 1993]. Figure 4c shows a vortex pair with the smaller ballistic swirl.

Moreover, a small modification of the initial conditions (Fig. 4d) projects the flow to a convergent spiral at the ground level and all vortices abort. Nonetheless, a gust can change these IC and one or more vortices occurs, refuting permanently the concept of *reliable* forecasts.

Besides, with $P_{5.a}$ set of parameters, the patterns can have the main features of the previous figure since the outcome of the simulation is a four-columnar dynamics with the persistent tiny ballistic swirl (fig. 5a). On the other hand, with the $P_{5.b}$ set of values, the dynamics can be utterly different: a vortex with a *serpentine shape* appears alone (fig. 5.b) and the motion converges to one of the two precedent stable solutions $S_1$ or $S_2$. This characteristic testifies the Sensitive Dependency on Parameters (S.D.P.).

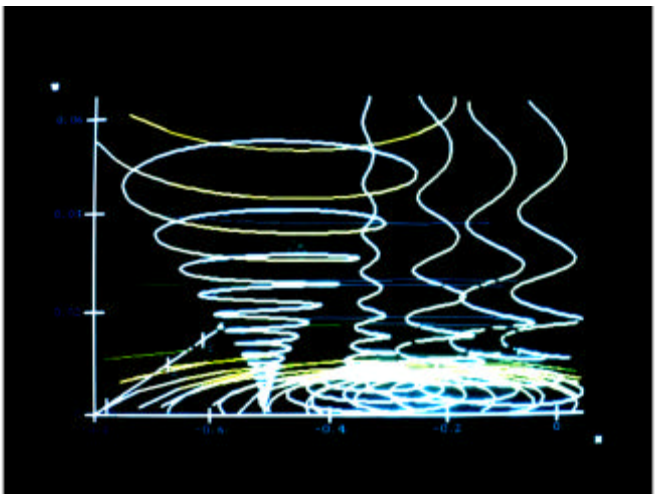
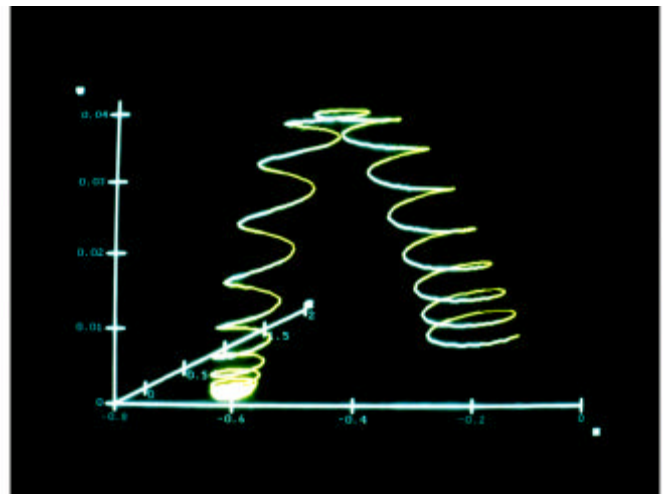

*(a) Four-column vortex dynamics with the ballistic swirl (left).*
$P_{5.a} (a, b, c, d, v, f, s) = ( 0.2, 0.6, 0.2, 10, 0.05, 1, 0.5 )$

*(b) Vortex with a serpentine shape.*
$P_{5.b} (a, b, c, d, v, f, s) = ( 0.2, 0.6, 0.1, 10, 0.5, 0.5, 0.5 )$

***Fig 5.** The Sensitive Dependency on Parameters (SDP)*
$IC(x_0, y_0, z_0, w_0 )= (0.1, 0.01, 0.01, 0)$

Short lifetime cycle, asymmetry and unpredictability are the highlights of this second class of vortices whose phase space volume is dissipative.

Finally, an additional advection linkage provides our dynamical system with an explanatory proof of tornadogenesis. By answering the following question: what is the outcome of distorted atmospheric rolls?, our research fills the gap between the 2D patterns and the 3D vortex structures.

Our heuristic dynamical system lays the foundation of an unified modelisation of vortices which associates theory and Direct Numerical Simulation in a new perspective.

## 5. Concluding remarks

Written to simulate the atmospheric vortices, our modelisation is based on a purely kinematic approach. For our analysis, the roll formula is simplified significantly since we do not need the explicit values of the fluid parameters. We select a modified Gause' oscillator as a first and sufficient step to emulate the 2D motion of a vertical eddy. We connect to this model of gyre a feedback loop in its orthogonal axis which constitutes the advection direction leading to a computational simulation and visualisation.

The several systems studied are derived exclusively from a single modelisation with basic and sophisticated twist processes since nullifying v disconnect the vertical shift and yields the first class of vortices. In a future paper, we will vary both s and v parameters for assesing the bifurcation diagram of the twist process and its transition regimes.

In an effort to investigate the vorticity mechanism we have carried out a variety of linkages. Our approach is heuristic seeing that fulfilling three embedded feedback into the model (a supplemental contortion of the twist process with an horizontal shift) did not yield new vortex patterns.

This rapid communication is a computed-assisted proof of occuring vortices from a twisted rotational fluid and provides a preliminary estimate of the general stability of the systems studied. However, the full model (sec. 4) deserves a deep analysis of its nonlinearity with special emphasis on the chaotic dynamics of short lifespan vortices since we detect SDIC and SDP.

Eventually, deriving our 2D model of gyre from Euler or Navier-Stokes equations is also a next step of this exploration but our theoretical choice is justified *since the numerically simulated* vortices *are realistic*!

Governed by a low dimensional system our perspective constitutes, to the best of our knowledge, the comprehensive modelisation of the kinematics of atmospheric vortices without violations of the fluid dynamics laws.

This research can be transposed to fields close to the atmospheric sciences, for example hydrodynamics when the roll is the Rayleigh-Bénard convection. Also this approach can be transposed to astrophysical sciences and enables the study of interstellar vortices when the gyre is created by a gravitational field.

This framework, which is just at its preliminary stage, can develop a new vortex experiments in order to establish a more definite relationship between the present theoretical perspective and experimental observations. Thus, this research will be provided with an unquestionable proof when *real* experiments create the patterns numerically depicted in a *new* apparatus (for example, wind tunnel forming a T with a rotating chamber...). We work for this plan.

## Appendix

Reduced systems of Navier-Stokes or Euler equations are often the central paradigm of rotational fluids studies. For the present modelisation, we introduce a dynamical equivalent as a convenient approximation of real aerologic motion of eddy or rotating thermal convection without loss of (kinematic) generality. However, the choice of a numerical analogue does not substitute the real equations governing the fluid dynamics.

Indeed, we choose to focus on one of the 2D oscillators, namely the Gause model (1934):

$$\begin{cases} dx/dt = x\,g(x) - y\,p(x) \\ dy/dt = -c\,y + y\,q(x) \end{cases}$$

describing the density of prey and predator species, i.e. x and y variables, in a confined biosystem. G, p and q are functions and c is a positive parameter. The Gause model is the generalized version [Ridane, 1988; Braun, 1993] of the Lotka-Volterra equation (1931), an eminent zoological oscillator where $g(x)= a$, $p(x) = bx$ and $q(x) = dx$ with a, b and d positive parameters. In comparison of this oscillator, we convert the signs and superimpose a supplemental nonlinearity, i.e. $q(x) = x^2$, to found our ODE model.

This singular methodology in the fluid dynamics science is chosen to emulate numerically the dynamical properties of the fully developed eddy flux by solving a very small set of equations.